 \documentclass[conference]{IEEEtran}
\usepackage{graphicx}
\usepackage{setspace}
\usepackage{amsmath}
\usepackage{amssymb} 
\usepackage{bm} 
\usepackage[normalem]{ulem}
\usepackage{cite}
\usepackage{float}
\usepackage[usenames,dvipsnames]{pstricks}
\usepackage{epsfig}
\usepackage{pst-grad} 
\usepackage{pst-plot} 
\usepackage{tabularx}
\usepackage[T1]{fontenc}
\newcommand{\lb} {\left}
\newcommand{\rb} {\right}
\newcommand{\nn} {\nonumber}

\allowdisplaybreaks
\usepackage{lipsum}
\usepackage{flushend}
\flushend

\usepackage{caption}
\usepackage{subcaption}
\usepackage[utf8]{inputenc}

\IEEEaftertitletext{\vspace{-2.4\baselineskip}}

\begin{document}
\onecolumn{\noindent © 2024. Personal use of this material is permitted. Permission from authors must be obtained for all other uses, in any current or future media, including reprinting/republishing this material for advertising or promotional purposes, creating new collective works, for resale or redistribution to servers or lists, or reuse of any copyrighted component of this work in other works.}
 \twocolumn{
\title{Secrecy Analysis of CSI Ratio-Based  Transmitter Selection with Unreliable Backhaul
   }

\author{
    \IEEEauthorblockN{Burhan Wafai$^{1}$, Ankit Dubey$^{2}$, 
    and Chinmoy Kundu$^{3}$   
    }
    \IEEEauthorblockA{$^{1}$ $^{2}$Department of EE, Indian Institute of Technology Jammu, Jammu \& Kashmir, India }
     \IEEEauthorblockA{$^{3}$Wireless Communications Laboratory, Tyndall National Institute, Dublin, Ireland}
    \textrm{\{$^{1}$burhan.wafai}, {$^{2}$ankit.dubey}\}@iitjammu.ac.in,{$^{3}$chinmoy.kundu@tyndall.ie},
     }

\maketitle
\thispagestyle{empty}
\pagestyle{empty}
\pagestyle{plain} 
\begin{abstract} 
This paper explores the secrecy performance of a multi-transmitter system with unreliable backhaul links.
To improve secrecy, we propose a novel transmitter selection (TS) scheme that selects a transmitter with the maximum ratio of the destination channel power gain to the eavesdropping channel power gain. 
The backhaul reliability factor is incorporated with the distribution of the channel power gain through the utilization of a mixture distribution. 
We evaluate the non-zero secrecy rate (NZR) and the secrecy outage probability (SOP) as well as their asymptotes in two scenarios of backhaul activity knowledge, where it is available and where it is unavailable.
The results illustrate that because of the unreliable backhaul, the proposed destination-to-eavesdropper channel power gain ratio-based TS scheme is constrained in terms of secrecy performance. However, performance enhancements are observed when the backhaul knowledge activity is utilized. Furthermore, the proposed scheme outperforms all the sub-optimal TS schemes and achieves nearly optimal performance without requiring noise power or the evaluation of the exact secrecy rate measurement.
\end{abstract}

\section{Introduction}
The Internet of Things (IoT) in the future wireless technologies will require high connectivity and data rates to accommodate a substantial volume of applications and devices \cite{fuqaha_iot}.
Future wireless technologies will be supported by heterogeneous small-cell networks to provide IoT services.  
In such heterogeneous networks, the backhaul establishes a connection between the access point and the small-cell base stations \cite{OnurBackhaul2022}. These backhaul links are either wireless or wired. Notably, wireless backhaul is becoming widely used rather than wired backhaul due to lower deployment costs \cite{Xiaohu20145G}. 

The wireless nature of IoT networks exposes them to the risk of eavesdropping. 
Physical layer security offers a solution to provide security to such networks \cite{wyner_wiretap}. 
Transmitter selection (TS) is one of the techniques for improving secrecy in multi-source wireless networks by increasing diversity gain of the legitimate channel without employing multiple antennas \cite{kim2016secrecy}. Depending on whether the channel state information (CSI) is available, TS schemes are broadly classified as optimal and sub-optimal selection schemes.
Optimal TS requires global CSI to select the transmitter with maximum secrecy rate, whereas the sub-optimal TS maximizes the legitimate channel rate, not requiring the eavesdropper channel CSI \cite{chinmoy_GC16,yang2013transmit, Vu2017Secure,Kundu15_dual, Kundu_TVT19,wafaiVTC}. 
Furthermore, the CSI of the backhaul significantly affects the system's performance \cite{wafaiVTC,chinmoy_letter2021,kotwalTVT_backhaul,wafai_IET}. The prior knowledge of the backhaul activity enables us to select the transmitter with an active backhaul link, whereas a transmitter corresponding to an inactive backhaul link may be selected because of the unavailability of backhaul activity knowledge.

\subsection{Related Works}
The security of the TS in a network with multiple transmitters has been extensively explored in literature \cite{chinmoy_GC16,yang2013transmit, Kundu_TVT19, wafaiVTC,  Vu2017Secure,chinmoy_letter2021, Kundu15_dual,kotwalTVT_backhaul,wafai_IET,kimBackhaulSelection }. 
Different TS schemes were investigated in \cite{Kundu_TVT19} for a system with a destination and an eavesdropper, with backhaul knowledge activity unavailable. Sub-optimal,  minimal eavesdropping, minimal interference, and optimal TS schemes were studied for a cognitive radio (CR) network with unreliable backhaul. 
The non-zero secrecy rate (NZR), secrecy outage probability (SOP), and ergodic secrecy rate (ESR) were derived. 
A similar system for a CR network was considered in \cite{wafaiVTC}, where the SOP of sub-optimal and optimal TS schemes were evaluated, with backhaul knowledge activity available.  
However,  the works in \cite{Kundu_TVT19} and \cite{wafaiVTC} 
 were primarily focused on the security of a CR network.

A sub-optimal and the optimal TS schemes were studied in \cite{kotwalTVT_backhaul} in a system consisting of a destination and multiple eavesdroppers. The SOP and ESR were derived under frequency selective fading for backhaul knowledge activity unavailable and available cases. 
The ESR of the optimal TS scheme was obtained for frequency flat fading channels with backhaul knowledge activity available in \cite{chinmoy_letter2021} for a system with a destination and multiple eavesdroppers.
Recently, a system consisting of a destination and multiple eavesdroppers was considered in \cite{wafai_IET} for frequency flat fading channels. A sub-optimal, minimum eavesdropping, and optimal TS schemes were studied for backhaul knowledge activity unavailable and available cases.

\subsection{Motivation and Novelty}
Although various TS schemes were explored in  \cite{chinmoy_letter2021,kotwalTVT_backhaul,wafai_IET}, none have specifically investigated a scheme that takes into account the maximum ratio of legitimate-to-eavesdropper channel CSI.
While the optimal TS scheme studied in \cite{ Kundu_TVT19,wafaiVTC,chinmoy_letter2021,kotwalTVT_backhaul,wafai_IET} involves the ratio of SNRs of the legitimate channel to the eavesdropping channel, the proposed TS scheme differs from the optimal TS scheme in the selection criteria. Unlike the optimal TS scheme, which aims to maximize the achievable secrecy rate, the proposed scheme focuses on maximizing the legitimate-to-eavesdropper channel power gains ratio to improve security.
Furthermore, the optimal TS scheme requires noise power of the receiving nodes (destination and eavesdropper) for the TS.
In contrast, the proposed scheme selects the transmitter based on the CSIs of the legitimate and eavesdropping channels without requiring noise power. In practical applications, measuring noise power can be challenging and resource-intensive. Therefore, a TS without the requirement of noise power simplifies the selection process and reduces complexity. Furthermore, we will illustrate that it also provides secrecy performance close to the performance of the optimal TS scheme.

Motivated by the above discussion, a wireless backhaul serving a small-cell network with a destination and an eavesdropper is considered where a novel channel CSI ratio-based transmitter selection (RTS) scheme is proposed that relies on the ratio of legitimate-to-eavesdropper channel power gains for the best transmitter selection without requiring the receiver noise power information.
To the best of the authors' knowledge, the RTS scheme has not been explored yet in the context of the secrecy of a multi-transmitter system.
{Despite the selection strategy, the system may still experience secrecy outage due to channel conditions and backhaul unreliability. 
Therefore, we evaluate the NZR and SOP for the RTS scheme incorporating the backhaul reliability factor into the analysis while ensuring that the transmitter selection is implemented. This makes the analysis challenging. 
The main contributions are:}
\begin{itemize}
    \item With the global CSI available, we introduce a novel CSI RTS scheme in which a transmitter with a maximum ratio of the legitimate link channel power gain to the eavesdropping link channel power gain is selected.
    \item The closed-form NZR and SOP expressions are obtained for the backhaul knowledge activity available and unavailable cases of the RTS scheme. We also evaluate the asymptotic NZR and SOP using the high SNR approximation to get better insights into the system.  
    \item We demonstrate that the proposed scheme performs better than all the sub-optimal schemes studied in \cite{wafai_IET}. Notably, it achieves performance comparable to the optimal TS scheme without requiring noise power or the evaluation of secrecy rate which simplifies the TS process in comparison to the optimal TS scheme.
    \item The results reveal that with the backhaul knowledge activity unavailable, employing additional transmitters does not lead to an improvement in performance. However, the resources can be judiciously used when the backhaul knowledge activity is available.
\end{itemize}

\section{System model}\label{section_system_model}
\begin{figure}
\centering 
\includegraphics[width=3.4in]{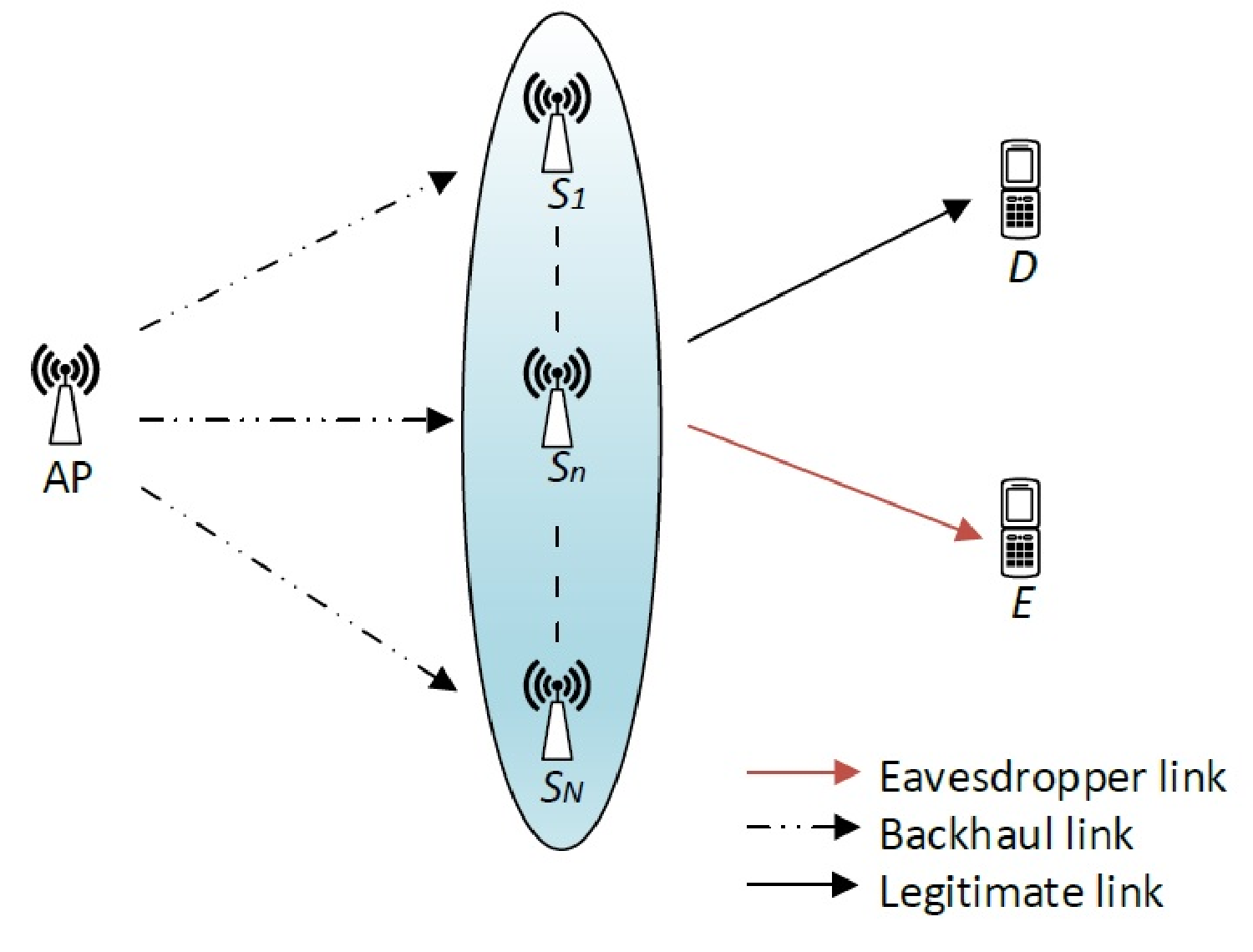} 
\caption{A small-cell network with backhaul links.}
\label{fig_SM2}
\end{figure}
A multi-transmitter system consisting of an access point (AP) serving  $K$  transmitter nodes $T_k$ where $k\in\{1, \ldots, K\}$  through the unreliable wireless backhaul links is considered, as depicted in Fig. \ref{fig_SM2}. $K$ transmitters communicate with a destination node $D$, and an eavesdropper $E$ intercepts the transmission, with each node having a single antenna.
We assume that the eavesdropper is a part of the network, and thus, its CSI is available \cite{PLS_relay}. 
All the links are assumed to be independent; specifically, we also assume that the channel gains of the links $T_k$-$D$, $h_{TD}^{(k)}$, for each $k\in\{1,\ldots, K\}$ are independent and identically distributed. Similarly, channel gains of the links $T_k$-$E$, $h_{TE}^{(k)}$,  for each $k\in\{1,\ldots,K\}$ are also assumed to be independent and identically distributed. However, the channel gains $h_{TD}^{(k)}$ and $h_{TE}^{(k)}$ for any $k$ are not necessarily identical.
Each link exhibits frequency flat Rayleigh fading and, therefore, the channel power gain $|h_{TX}^{(k)}|^2$ where $X\in\{D, E\}$ is exponentially distributed with parameter $\lambda_{X}$.
The CDF and PDF  of the channel power gain $|h_{TX}^{(k)}|^2$ is 
\begin{align}
\label{Cdf_SX_without_knowledge}
F_{|h_{TX}^{(k)}|^2}(x)&=1-\exp(-\lambda_{X} x),\\
\label{pdf_SX_without_knowledge}
f_{|h_{TX}^{(k)}|^2}(x)&=\lambda_{X}\exp(-\lambda_{X} x), 
\end{align}
respectively. 
Each backhaul link serving a transmitter has a probability of link failure. 
The reliability of the backhaul link is modeled by an independent Bernoulli random variable with success probability $\Delta$ and failure probability $1-\Delta$. 
The end-to-end link distribution of AP-$T_k$-$D$ link where $k\in\{1,\ldots, K\}$ is expressed with the help of a mixture distribution \cite{Kundu_TVT19}, which incorporates the backhaul reliability factor $\Delta$. 
A mixture distribution is used to express the PDF of the link SNR from AP to $X$ via $T_k$ is expressed as 
\cite{chinmoy_letter2021}
\begin{align}\label{mixture_distribution}
f_{|\hat{h}_{TX}^{(k)}|^2}(x)=(1-\Delta)\delta(x)+\Delta f_{|h_{TX}^{(k)}|^2}(x)
\end{align}
where $\delta(x)$ is the Dirac delta function, $\hat{h}_{TX}^{(k)}$ represents the channel that includes the backhaul reliability factor.
The  corresponding CDF will be 
\begin{align}\label{cdf_SD_with_knowledge}
F_{|\hat{h}_{TX}^{(k)}|^2}(x)&=(1-\Delta)u(x)+ \Delta\lb(1-e^{-\lambda_{D} x}\rb),
\end{align}
where $u(x)$ is the unit step function \cite{wafaiVTC}.
Additive white Gaussian noise (AWGN) is assumed to be added at each receiver node with zero mean and variance $\sigma_X$ where $X\in\{D, E\}$.
We study the RTS scheme in two cases: backhaul knowledge activity available and backhaul knowledge activity unavailable.

\subsection{Backhaul Activity Knowledge  Available}
With the backhaul knowledge activity known, the transmitter is chosen among those for which the backhaul links are operational. We select the transmitter ($k^*$) for which the ratio of legitimate-to-eavesdropper channel power gain is maximum.
Assuming $\mathcal{A}^{(q)}$ is the subset of transmitters consisting of $q$ active backhaul links, i.e., $\mathcal{A}^{(q)}\subseteq\{1,\ldots, K\}$ and $\mathcal{A}^{(q)} \ne \emptyset$, the ratio of $T_{k^*}$-$D$ and $T_{k^*}$-$E$ channel power gains for the selected transmitter $k^*$  is 
\begin{align}\label{max_ratio_with}
\hat{\Gamma}^{(k^*)}=\max_{k\in\mathcal{A}^{(q)}}  \lb(\frac{|\hat{h}_{TD}^{(k)}|}{|\hat{h}_{TE}^{(k)}|}\rb)^2.
\end{align} 
Since the destination and eavesdropper link have a common backhaul link, the backhaul reliability factor has to be incorporated in either the $T_k$-$D$ or $T_k$-$E$ link. Here, we include it in the $T_k$-$D$ link.
Using the CDF definition, the distribution of $\hat{\Gamma}^{(k^*)}$  is given by
\begin{align}\label{cdf_ratio_with}
F_{\hat{\Gamma}^{(k^*)}}(x)&=\mathbb{P}\lb[\max_{k\in\mathcal{A}^{(q)}} \Big\{\frac{|\hat{h}_{TD}^{(k)}|^2}{|\hat{h}_{TE}^{(k)}|^2}\Big\}\le x\rb]\nn\\
&=\lb(\int_{0}^{\infty}F_{|\hat{h}_{TD}^{(k)}|^2}(xy)f_{|\hat{h}_{TE}^{(k)}|^2}(y)dy\rb)^{K}\nn\\
&=(1-\Delta)^{K}u(x)\nn\\
&+\sum_{k=1}^{K}\binom{K}{k}(-1)^{k+1}\Delta^k\lb(1-\frac{\lambda_{E}}{\lambda_{D} x+\lambda_{E}}\rb)^{k},
\end{align}
where the distributions of  $f_{|\hat{h}_{TE}^{(k)}|^2}(\cdot)$ and $F_{|\hat{h}_{TD}^{(k)}|^2}(\cdot)$  are used from (\ref{pdf_SX_without_knowledge}) and (\ref{cdf_SD_with_knowledge}), respectively. 
 The corresponding PDF is
\begin{align}\label{pdf_ratio_with}
f_{\hat{\Gamma}^{(k^*)}}(x)&=(1-\Delta)^{K}\delta(x)\nn\\
&+\sum_{k=1}^{K}\binom{K}{k}(-1)^{k+1}\Delta^k\lb(\frac{\lambda_{E}}{\lambda_{D} x+\lambda_{E}}\rb)^{k}.
\end{align}

\subsection{Backhaul Activity Knowledge  Unavailable}
In this case, the TS is carried out without taking into account whether the backhaul link is active or inactive. The transmitter $k^*$  which has the channel power gain ratio $\Gamma^{(k^*)}$  maximum is selected without the backhaul knowledge activity. Thus, 
\begin{align}
\Gamma^{(k^*)}=\max_{k\in\{1,\ldots, K\}}  \lb(\frac{|h_{TD}^{(k)}|}{|h_{TE}^{(k)}|}\rb)^2. 
\end{align}
It is important to highlight that the selected $k^*$ link may be inactive.
The CDF of $\Gamma^{(k^*)}$ is expressed as
\begin{align}\label{cdf_ratio_without}
F_{\Gamma^{(k^*)}}(x)&=\mathbb{P}\lb[\max_{k\in\{1,\ldots, K\}} \Big\{\frac{|h_{TD}^{(k)}|^2}{|h_{TE}^{(k)}|^2}\Big\}\le x\rb]\nn\\
&=1-\sum_{k=1}^{K}\binom{K}{k}(-1)^{k+1}\lb(\frac{\lambda_{E}}{\lambda_{D} x+\lambda_{E}}\rb)^{k}.
\end{align}
The corresponding PDF is written as
\begin{align}\label{pdf_ratio_without}
f_{\Gamma^{(k^*)}}(x)&=
\sum_{k=1}^{K}\binom{K}{k}(-1)^{k+1}\frac{k\lambda_{E}^k\lambda_{D}}{\lb(\lambda_{D} x+\lambda_{E}\rb)^{k+1}}.
\end{align}
We incorporate the backhaul reliability factor to get the end-to-end PDF of the 
AP to $D$ link following (\ref{mixture_distribution})  as
\begin{align}\label{pdf_ratio_without1}
f_{\hat{\Gamma}^{(k^*)}}(x)&= (1-\Delta)\delta (x)+\Delta f_{\Gamma^{(k^*)}}(x). 
\end{align}
Next, we obtain the NZR and SOP for available and unavailable cases of backhaul knowledge activity using the distributions derived in this section.

\section{Non-zero Secrecy Rate}\label{section_nzsr}
The NZR of the RTS scheme for both the backhaul knowledge activity available and unavailable cases is evaluated in this section. In this regard, we first define the secrecy rate corresponding to any transmitter $T_{k}$ as
\begin{align}\label{secrecy_capacity}
  C_S^{(k)}=\max\Bigg\{\log_2\lb(\frac{1+\frac{|\hat{h}_{TD}^{(k)}|^2}{\sigma_D}}{1+\frac{|\hat{h}_{TE}^{(k)}|^2}{\sigma_E}}\rb),0\Bigg\},  
\end{align}
where $\frac{|\hat{h}_{TX}^{(k)}|^2}{\sigma_X}$ represents the SNR of the link $T_{k}$-$X$ where $X\in\{D,E\}.$
Consequently, the NZR of the RTS scheme is the probability that the secrecy rate of the selected transmitter $C_S^{(k^*)}$ exceeds zero.
\subsection{Backhaul Activity Knowledge  Available}\label{subsection_nzsr_with}
The availability of backhaul knowledge activity enables us to select the transmitter among the active links. 
Let us assume there exist $q$ backhaul links that are active, i.e., the cardinality of the set  $\mathcal{A}^{(q)}$ is $|\mathcal{A}^{(q)}|=q$.
Since independent and identically distributed Bernoulli random variables govern the activity of the backhaul links,
the NZR is obtained by summing up all the possible combinations of $q$ active links as 
\begin{align}\label{nz_rts_with}
&\mathcal{P}_{NZ}=1-\mathcal{P}_Z=1-\sum_{q=1}^{K}\binom{K}{q}\mathcal{P}_{Z}^{(q)},
\end{align}
where
$\mathcal{P}_{Z}$ is the probability that $C_S^{(k^*)}$ is less than or equal to zero. $\mathcal{P}_{Z}$ is the sum of all probabilities $\mathcal{P}_{Z}^{(q)}$ that the secrecy rate $C_S^{(k^*)}$ is less than or equal to zero when there exist $q$ active backhaul links. The term $\binom{K}{q}$  in the equation signifies the number of possible ways $q$ backhauls can be active. $\mathcal{P}_{Z}^{(q)}$ takes $\Delta$ into account because of the consideration of mixture distribution in (\ref{mixture_distribution}). It is evaluated first by finding the probability that $C_{S}^{(k)}$ is less than or equal to zero for each link $k\in\mathcal{A}^{(q)}$ provided that the link has the maximum channel power gain ratio, i.e., the channel power gain ratio is greater than the maximum of the rest of the corresponding channel power gain ratios. Then, by following the law of total probability, $\mathcal{P}_{Z}^{(q)}$ is obtained by summing all the above probabilities corresponding to each link in $\mathcal{A}^{(q)}$ as 
\begin{align}
\label{eq_zero_with}
&\mathcal{P}_{Z}^{(q)}=\sum\limits_{k\in\mathcal{A}^{(q)}} \mathbb P\lb[C_{S}^{(k)}\le 0\Bigg{|} \frac{|\hat{h}_{TD}^{(k)}|^2}{|\hat{h}_{TE}^{(k)}|^2}>\bar{\Gamma}^{(k)}\rb]\\
&=\sum\limits_{k\in\mathcal{A}^{(q)}}\mathbb P\lb[\frac{|\hat{h}_{TD}^{(k)}|^2}{\sigma_D}<\frac{|\hat{h}_{TE}^{(k)}|^2}{\sigma_E}\Bigg{|} |\hat{h}_{TD}^{(k)}|^2>|\hat{h}_{TE}^{(k)}|^2\bar{\Gamma}^{(k)}\rb]\nn \\
\label{eq_prob_zero}
&=\sum\limits_{k\in\mathcal{A}^{(q)}}\frac{\sigma_D}{\sigma_E} \int_{0}^{\infty}\int_{0}^{1}\int_{yz}^{y\sigma_D/\sigma_E}\nn\\
& f_{|\hat{h}_{TD}^{(k)}|^2}(x\sigma_D/\sigma_E) f_{\bar{\Gamma}^{(k)}}(z) f_{|\hat{h}_{TE}^{(k)}|^2}(y)dx dz dy,
\end{align}
where {$\bar{\Gamma}^{(k)}=\max_{\substack{n\in\mathcal{A}^{(q)} \\ n \ne k}} \Big\{\frac{|\hat{h}_{TD}^{(n)}|^2}{|\hat{h}_{TE}^{(n)}|^2}\Big\}$.}
The distribution of $\bar{\Gamma}^{(k)}$ is easily derived following (\ref{cdf_ratio_with}). Consequently, the PDF of $\bar{\Gamma}^{(k)}$ will be same as (\ref{pdf_ratio_with}), with  $K$ replaced by  $K-1$.
Using (\ref{pdf_SX_without_knowledge}), (\ref{mixture_distribution}), (\ref{pdf_ratio_with}), (\ref{nz_rts_with}), and (\ref{eq_prob_zero}) the NZR is finally written as
\begin{align}\label{eq_nzsr_with}
&\mathcal{P}_{NZ}
=1-\Bigg[(1-\Delta)^K+\frac{K\Delta(1-\Delta)^{K-1}\sigma_{D}{\lambda}_{D}}{\sigma_{E}{\lambda}_{E}+\sigma_{D}{\lambda}_{D}}+\sum_{q=1}^{K}\sum_{n=1}^{q-1}\nn\\
& \binom{K}{q}\binom{q-1}{n}\frac{(-1)^{n+1}nq\Delta^{n+1}\sigma_{E}{\lambda}_{E}}{(\sigma_{E}{\lambda}_{E}+\sigma_{D}{\lambda}_{D})^{n+1}}\Bigg(-\frac{1}{n}\Big(-\sigma_{E}^n{\lambda}_{E}^n\nn\\
&+(\sigma_{E}{\lambda}_{E}+\sigma_{D}{\lambda}_{D})^n  \Big)+\lb(\frac{(\sigma_{E}{\lambda}_{E}+\sigma_{D}{\lambda}_{D})^{n+1}}{\sigma_{E}{\lambda}_{E}}-\sigma_{E}^n{\lambda}_{E}^n\rb)\nn\\
&\times \Bigg(\sum_{i=1}^{n}(-1)^{n-i}\frac{(n-i)!(i-1)!}{n!}+\frac{(-1)^n}{(n+1)}\Bigg)\Bigg)\Bigg].
\end{align}
\subsection{Backhaul Activity Knowledge  Unavailable}\label{subsection_nzsr_without}
The NZR for this case is evaluated without prior knowledge of backhaul link activity. As each link has a failure probability $(1-\Delta)$, the NZR is expressed as
\begin{align}\label{nz_rts_without}
\mathcal{P}_{NZ}&=(1-\Delta)\times 0+\Delta\lb(1-\mathcal{P}_{Z}  \rb),
\end{align}
where  $\mathcal{P}_{Z}$ is the probability that $C_S^{(k^*)}$ is less than or equal to zero as if the backhaul link is active for selected transmitter $T_{k^*}$, i.e., assuming channel power gain as $|h_{TX}^{(k^*)}|^2$ in $C_S^{(k^*)}$ without incorporating the backhaul in (\ref{secrecy_capacity}). $\mathcal{P}_{Z}$ is evaluated following the similar logic in (\ref{eq_prob_zero}) as
\begin{align}
\label{eq_zero_without}
\mathcal{P}_{Z}=&\sum_{k=1}^{K}\mathbb P\lb[C_{S}^{(k)}\le 0\Bigg{|} \frac{|{h}_{TD}^{(k)}|^2}{|{h}_{TE}^{(k)}|^2}>  \bar{\Gamma}^{(k)} \rb] \\
\label{eq_prob_zero_without}
=&\frac{K\sigma_D}{\sigma_E} \int_{0}^{\infty}\int_{0}^{1}\int_{yz}^{y\sigma_D/\sigma_E}\nn\\
&f_{|{h}_{TD}^{(k)}|^2}(x\sigma_D) f_{\bar{\Gamma}^{(k)}}(z) f_{|{h}_{TE}^{(k)}|^2}(y\sigma_E)dx dz dy,
\end{align}
where
{$\bar{\Gamma}^{(k)}=\max_{\substack{n\in\{1,\ldots,K\} \\ n \ne k}} \Big\{\frac{|h_{TD}^{(n)}|^2}{|h_{TE}^{(n)}|^2}\Big\}$}. In contrast to (\ref{eq_zero_with}), the summation in (\ref{eq_zero_without}) is over $K$ as the backhaul activity knowledge is not available. Due to the independent and identical distributions of $C_{S}^{(k)}$ for any $k$ in (\ref{eq_zero_without}), the probabilities in (\ref{eq_zero_without}) for each $k$ is the same. Hence, $K$ is multiplied with the probability corresponding to any $k$ in (\ref{eq_prob_zero_without}). Using (\ref{mixture_distribution}), (\ref{pdf_SX_without_knowledge}), (\ref{pdf_ratio_without}),  (\ref{nz_rts_without}), and (\ref{eq_prob_zero_without}) we get
\begin{align}\label{eq_nzsr_without}
&\mathcal{P}_{NZ}
=\Delta\Bigg[1-K\sum_{n=1}^{K-1}\binom{K-1}{n}\frac{(-1)^{n+1}n\sigma_{E}{\lambda}_{E}}{(\sigma_{D}{\lambda}_{D}+\sigma_{E}{\lambda}_{E})^{n+1}}\nn\\
&\times\Bigg(\frac{\sigma_{E}^n{\lambda}_{E}^n-(\sigma_{E}{\lambda}_{E}+\sigma_{D}{\lambda}_{D})^n}{n} +\Big(\frac{(\sigma_{E}{\lambda}_{E}+\sigma_{D}{\lambda}_{D})^{n+1}}{\sigma_{E}{\lambda}_{E}}\nn\\
&-\sigma_{E}^n{\lambda}_{E}^n\Big) \lb(\sum_{i=1}^{n}(-1)^{n-i}\frac{(n-i)!(i-1)!}{n!}+\frac{(-1)^n}{(n+1)}\rb)\Bigg)\Bigg].
\end{align}
It is noticed from (\ref{eq_nzsr_without}) that the NZR is proportional to $\Delta$. This signifies that the secrecy is enhanced with $\Delta$. We also note that the NZR increases with $K$ due to the summation of $K-1$ terms in the second term of (\ref{eq_nzsr_without}). 

\section{Secrecy Outage Probability} \label{section_sop}
The SOP is the probability that the achievable secrecy rate of the system falls below a threshold secrecy rate $R_{th}$ \cite{wang2015security}. The SOP of the RTS scheme is evaluated for both cases of backhaul knowledge activity.
\subsection{Backhaul Activity Knowledge  Available}\label{subsection_sop_with}
The SOP is obtained using the same procedure considered in (\ref{nz_rts_with}) and (\ref{eq_zero_with}) for the NZR. It is derived by evaluating the probability corresponding to each link that the secrecy rate of that link falls below the threshold $R_{th}$ provided the channel power gain ratio of that link is maximum. Then, due to the law of total probability, we sum over all such probabilities corresponding to each link in set $\mathcal{A}^{(q)}$ and also sum over all possibilities of $q$ active links. Thus, the SOP is expressed as
\begin{align}\label{sop_rts_with}
&\mathcal{{P}}_{out}=\sum_{q=1}^{K}\binom{K}{q}\sum\limits_{k\in\mathcal{A}^{(q)}}\mathbb{P}\lb[C_{S}^{(k)}<{R_{th}}\bigg{|} \frac{|\hat{h}^{(k)}_{TD}|^2}{|\hat{h}^{(k)}_{TE}|^2}>\bar{\Gamma}^{(k)}\rb]\nn\\
&=\sum_{q=1}^{K}\binom{K}{q}\sum\limits_{k\in\mathcal{A}^{(q)}}\nn\\
&\mathbb{P}\lb[\frac{|\hat{h}^{(k)}_{TD}|^2}{\sigma_D}<\rho\frac{|\hat{h}^{(k)}_{TE}|^2}{\sigma_E}+\rho-1\Bigg{|} \frac{|\hat{h}^{(k)}_{TD}|^2}{|\hat{h}^{(k)}_{TE}|^2}>\bar{\Gamma}^{(k)}\rb]\nn\\
&=\sum_{q=1}^{K}\binom{K}{q}{q}\int_{0}^{\infty}\int_{0}^{\chi(x)/{y}}\int_{yz}^{\chi(x)}\nn\\
& f_{|\hat{h}_{TD}^{(k)}|^2}\lb(\chi(x)\rb) f_{\bar{\Gamma}^{(k)}}(z) f_{|\hat{h}_{TE}^{(k)}|^2}(y)dx dz dy,
\end{align}
where $\bar{\Gamma}^{(k)}=\max_{\substack{n\in\mathcal{A}^{(q)} \\ n \ne k}} \Big\{\frac{|\hat{h}_{TD}^{(n)}|^2}{|\hat{h}_{TE}^{(n)}|^2}\Big\}$,  $\rho=2^{R_{th}}$ and $\chi(x)=(\rho x+\sigma_E(\rho-1))\frac{\sigma_D}{\sigma_E}$. Finally, on substituting (\ref{pdf_SX_without_knowledge}), (\ref{mixture_distribution}), and (\ref{pdf_ratio_with})   into (\ref{sop_rts_with}) we get the closed-form expression as 
\begin{align}\label{eq_sop_with}
&\mathcal{P}_{out}=(1-\Delta)^K+\Delta(1-\Delta)^{K-1}K\Big(\frac{-\sigma_{E}\lambda_{E} e^{-\sigma_{D}\lambda_{D}(\rho-1)}}{\rho\sigma_{D}\lambda_{D}+\sigma_{E}\lambda_{E}}\nn\\
&+1\Big)+\sum_{q=2}^{K}\binom{K}{q}q\Bigg[\sum_{n=1}^{K-1}\binom{K-1}{n}(-1)^{n+1}\Delta^{n+1}n\Bigg[\sum_{r=1}^{n}\nn\\
&\frac{(-1)^{n-r}}{{n!}} (r-1)!(n-r)!+\frac{(-1)^n\gamma(n+1)}{(n+1)}-\frac{\sigma_{D}{\lambda}_{D}^{n+1}}{a^{n+1}n!}\sum_{r=1}^{n}\nn\\
&(r-1)!(-1)^{n-r}\sum_{i=0}^{n}\binom{n}{i}(-b)^{n-i}\Gamma(i-r+1,b)+\frac{(-1)^n}{a^{n+1}}\nn\\
&+\frac{(\sigma_{E}{\lambda}_{E})^{n+1}}{(n+1)!}\Big((-b)^{n+1}\text{Ei}(-b)-e^{-b}\sum_{l=0}^{n}(n-l)!(-b)^l\Big)\nn\\
&+\frac{(\sigma_{E}{\lambda}_{E})^{n+1}}{na^{n+1}}\sum_{p=0}^{n}\binom{n}{p}\frac{\Gamma(p-n+1,b)}{a(-b)^{p-n}}-\frac{e^{-b}}{na\sigma_{E}{\lambda}_{E} }\Bigg],
\end{align}
where $\gamma(\cdot)$ is the lower gamma integral, $\Gamma(\cdot)$ is the upper gamma integral, $\text{Ei}(\cdot)$ is the exponential integral, $a=\rho\sigma_{D}{\lambda}_{D}+\sigma_{D}{\lambda}_{D}$, and $b=\sigma_{D}{\lambda}_{D}(\rho-1)$.

\subsection{Backhaul Activity Knowledge  Unavailable}\label{subsection_sop_without}
The SOP is obtained in this section following the approach used for the NZR in (\ref{nz_rts_without}). Therefore, by following the similar steps as in (\ref{eq_zero_without}) and (\ref{eq_prob_zero_without}), and using   (\ref{pdf_SX_without_knowledge}), (\ref{mixture_distribution}), (\ref{pdf_ratio_without}), and (\ref{eq_sop_without1}), the SOP is expressed  as  
\begin{align}\label{eq_sop_without1}
&\mathcal{P}_{out}=(1-\Delta)+ \Delta\mathbb{P}\lb[C_{S}^{(k)}<{R_{th}}\bigg{|} \frac{|h^{(k)}_{TD}|^2}{|h^{(k)}_{TE}|^2}>\bar{\Gamma}^{(k)}\rb]\nn \\
&=(1-\Delta)+\Delta K \sum_{n=1}^{K-1}\binom{K-1}{n}(-1)^{n+1}n\Bigg[\sum_{r=1}^{n}\frac{(-1)^{n-r}}{n!}\nn\\
&\times (r-1)!(n-r)!+\frac{(-1)^n\gamma(n+1)}{(n+1)}-\frac{(\sigma_{E}{\lambda}_{E})^{n+1}}{a^{n+1}n!}\sum_{r=1}^{n}\nn\\
&(r-1)!(-1)^{n-r}\sum_{i=0}^{n}\binom{n}{i}(-b)^{n-i}\Gamma(i-r+1,b)+\frac{(-1)^n}{a^{n+1}}\nn\\
&\times\frac{(\sigma_{E}{\lambda}_{E})^{n+1}}{(n+1)!}\Big((-b)^{n+1}\text{Ei}(-b)-e^{-b}\sum_{l=0}^{n}(n-l)!(-b)^l\Big)\nn\\
&+\frac{(\sigma_{E}{\lambda}_{E})^{n+1}}{na^{n+1}}\sum_{p=0}^{n}\binom{n}{p}\frac{\Gamma(p-n+1,b)}{a(-b)^{p-n}}-\frac{e^{-b}}{na\sigma_{E}{\lambda}_{E} }\Bigg].
\end{align}
It is difficult to get insights from the closed-form equations derived in Section \ref{section_nzsr}  and \ref{section_sop} because of the complex nature of the equations in (\ref{eq_nzsr_with}), (\ref{eq_nzsr_without}), (\ref{eq_sop_with}) and (\ref{eq_sop_without1}). Therefore, we determine the asymptotic NZR and SOP in the next section.

{We note that the noise power information is needed for the performance analysis (NZR and SOP); however,  the system does not require this information for the implementation of the proposed RTS scheme.}

\section{Asymptotic Analysis}
The approximation of high SNR is utilized in this section to find the asymptotic NZR and SOP for both the backhaul knowledge activity cases. 
We use the approximation of high SNR, i.e., $1/\lambda_D\rightarrow\infty$ or $\lambda_D\rightarrow 0$ in the derived closed-form expressions to get insights and the effect of $\Delta$ and $K$ on the system's performance.
\subsection{Asymptotic NZR}
\subsubsection{Backhaul Activity Knowledge  Available}
Using the high-SNR approximation in (\ref{eq_nzsr_with}), we notice that all the terms approach zero except the first term inside the square bracket. Hence, the asymptotic NZR for the RTS scheme with backhaul knowledge activity available is $1-(1-\Delta)^K$. The asymptotic NZR depends on $\Delta$ and $K$, illustrating that the availability of backhaul knowledge activity enables the use of resources (number of transmitters) judiciously, thereby enhancing the secrecy performance.
\subsubsection{Backhaul Activity Knowledge  Unavailable}
Similarly, we find that all the terms approach zero as $1/\lambda_D\rightarrow\infty$ in (\ref{eq_nzsr_without}), except the first term, and thus the asymptotic NZR will be equal to $\Delta$. 
This implies that the NZR becomes independent of all parameters at high SNRs, depending only on $\Delta$.  

\subsection{Asymptotic SOP}
\subsubsection{Backhaul Activity Knowledge  Available}
Using the approximation $1/\lambda_D\rightarrow\infty$ in (\ref{eq_sop_with}), all the terms, except the first one, tend to zero. Consequently, the asymptotic SOP when backhaul knowledge activity is available is $(1-\Delta)^K$.
This observation is similar to the NZR knowledge available case where the asymptote depends on $\Delta$ and $K$. 
\subsubsection{Backhaul Activity Knowledge  Unavailable}
By following a similar procedure of $1/\lambda_D\rightarrow\infty$ in (\ref{eq_sop_without1}), we notice that the asymptotic SOP is $1-\Delta$. Thus, when backhaul knowledge activity is unavailable, the asymptotic SOP  depends only on $\Delta$. A similar observation was found in the NZR knowledge unavailable case.
\section{Numerical Results}
This section presents and discusses the analytical results of the RTS scheme. We validate the mathematical equations by Monte Carlo simulations, and it is evident from the figures that the analytical results perfectly match the simulation.
The parameters are set as $R_{th}=1$ bits/s/Hz, $1/\lambda_{E}=8$ dB, $\sigma_D=1$ dB and $\sigma_E=10$ dB.
The results are presented for backhaul knowledge available and when it is available.

\begin{figure}
    \centering 
    \includegraphics[width=3.4in]{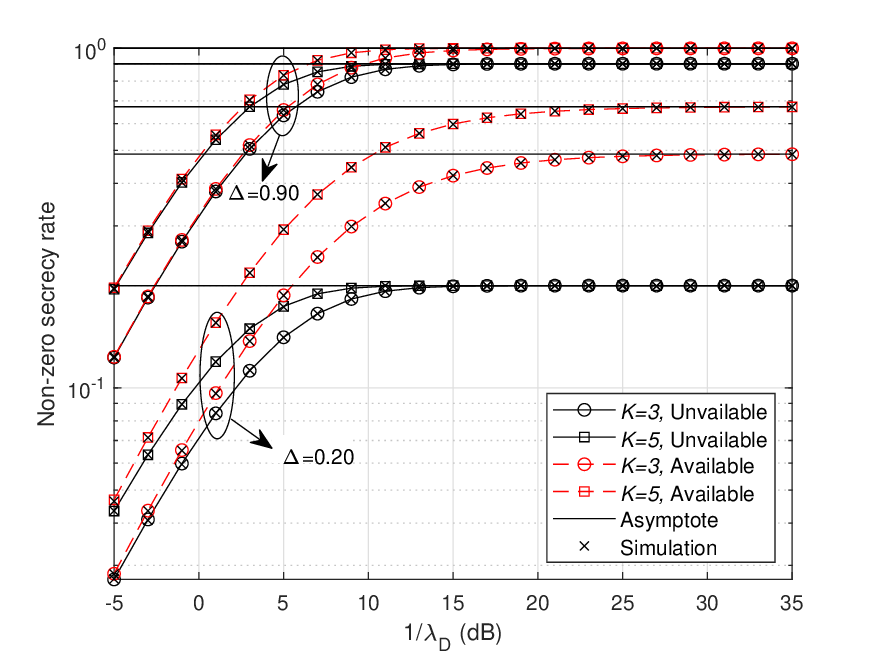}
    \caption{The NZR versus $1/\lambda_{D}$ varying $K$  and $\Delta$.}
    \label{fig_nzsr}
\end{figure}
Fig. \ref{fig_nzsr} presents the NZR versus the SNR $1/\lambda_{D}$(dB) by changing $K=\{3,5\}$ and $\Delta=\{0.9, 0.2\}$. 
We observe that with an increase in $K$ and $\Delta$, the secrecy is enhanced. We also notice that the NZR saturates to a constant value at high SNR. 
We notice the backhaul knowledge activity has a significant impact on secrecy performance and the effect of backhaul knowledge activity is more significant when the backhaul is less reliable, i.e., $\Delta$ = $0.2$. 
This shows that with prior knowledge of backhaul link activity, the unreliability of backhaul can be compensated.

\begin{figure}
    \centering 
    \includegraphics[width=3.4in]{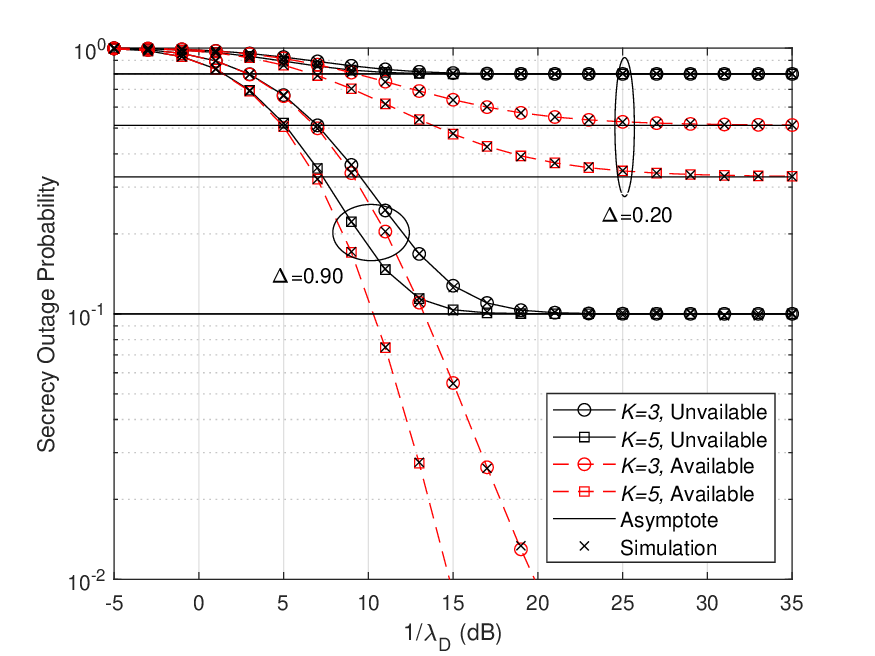}
    \caption{SOP versus $1/\lambda_{D}$ varying $K$  and $\Delta$.}
    \label{fig_sop}
\end{figure}
In Fig. \ref{fig_sop}, the SOP is plotted against the SNR $1/\lambda_{D}$(dB) for different  $K$ and  $\Delta$. The same trend is found for the SOP performance in this figure as the NZR. 
We notice that the backhaul knowledge activity available case outperforms the cases in which it is unavailable.
Furthermore, the SOP reaches a constant value at high SNR, which decreases with $\Delta$. 
The saturation levels of the knowledge available cases are lower than the knowledge unavailable cases.
This signifies that the transmitters can be well utilized when backhaul knowledge activity is available. 

\begin{figure}
    \centering 
    \includegraphics[width=3.4in]{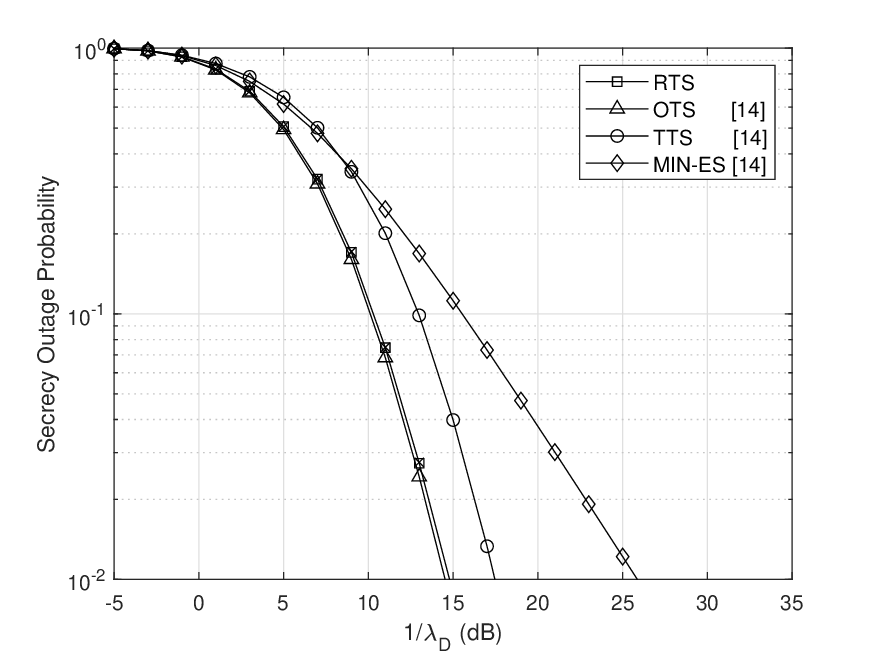}
    \caption{SOP comparison of different transmitter selection schemes.}
    \label{fig_sp}
\end{figure}
Fig. \ref{fig_sp} shows a comparison of the RTS scheme and various TS schemes from \cite{wafai_IET} when backhaul knowledge activity is available with $K=5$ and $\Delta=0.9$.  
The comparison includes traditional transmitter selection (TTS), which maximizes legitimate channel rate; minimal eavesdropping selection (MIN-ES), in which the transmitter with the worst eavesdropping channel is selected; and the optimal TS scheme.
We observe that the RTS scheme outperforms the TTS and MIN-ES schemes while closely following the optimal TS scheme from \cite{wafai_IET}. 
This signifies that achieving optimal performance can be achieved without considering noise power or secrecy rate. Thus simplifying the selection process.

\section{Conclusion}
A destination-to-eavesdropper channel CSI ratio-based transmitter selection scheme has been proposed to improve the secrecy of the system. The NZR and SOP have been evaluated in closed form along with their asymptotes for the RTS scheme for the backhaul knowledge activity available and unavailable cases. 
The results show that the secrecy performance increases when the number of transmitters increases. The secrecy performance also increases with the backhaul reliability factor. In the high SNR regime, we observe that when the backhaul knowledge activity is unavailable, both the NZR and the SOP saturate to a constant value that depends only on the backhaul reliability factor. The prior knowledge of backhaul activity enables more efficient utilization of resources.
We demonstrate that the RTS scheme outperforms all the sub-optimal TS schemes and achieves nearly optimal performance without requiring noise power or the evaluation of the exact secrecy rate measurement.
\section{Acknowlwdgement}
This work was supported in part by the Science Foundation Ireland (SFI) under Grant Number 22/IRDIFA/10425, Tata Consultancy Services (TCS) Foundation through its TCS Research Scholar Program,  and the Ministry of Electronics and Information Technology (MeitY), Govt. of India, through its project on capacity building for human resource development in unmanned aircraft systems (drone and related technology) with Ref. No. L14011/29/2021-HRD.

\bibliographystyle{IEEEtran}
\bibliography{IEEEabrv,RTS}
\end{document}